\documentclass[prb,twocolumn,showpacs,amsmath,amssymb]{revtex4}

\usepackage{graphicx}% Include figure files
\usepackage{dcolumn}% Align table columns on decimal point
\usepackage{bm}% bold math
\begin{document}
\newcommand{\jt}{$(h\pm\delta,k\pm\delta,l)$}
\newcommand{\co}{$(h\pm2\delta,k\pm2\delta,l)$}
\newcommand{\bilayer}{La$_{2-2x}$Sr$_{1+2x}$Mn${_2}$O$_{7}$}
\preprint{PRB/version2.0/TAWB}

\title{Orbital Bi-Stripes in High Doped Bilayer Manganites}

\author{T.A.W. Beale}
\author{P.D. Spencer}
\author{P.D. Hatton}
\email{p.d.hatton@dur.ac.uk} \affiliation{Department of Physics,
University of Durham, Rochester Buildings, South Road, Durham, DH1
3LE, United Kingdom.}
\author{S.B. Wilkins} \affiliation{European
  Commission, Joint Research Center, Institute for Transuranium Elements,
Hermann von Helmholtz-Platz 1, 76344 Eggenstein-Leopoldshafen,
  Germany}\affiliation{European Synchrotron Radiation
  Facility, Bo\^\i te Postal 220, F-38043 Grenoble Cedex, France}
\author{M. v. Zimmermann}
\affiliation{Hamburger Synchrotronstrahlungslabor (HASYLAB) at
Deutsches Elektronen-Synchrotron (DESY), Notkestra{\ss}e 85,
D-22603 Hamburg, Germany.}
\author{S.D. Brown}
\affiliation{European Synchrotron Radiation
  Facility, Bo\^\i te Postal 220, F-38043 Grenoble Cedex, France}
\author{D. Prabhakaran}
\author{A.T. Boothroyd}
\affiliation{Department of Physics, University of Oxford,
Clarendon Laboratory, Parks Road, Oxford, OX1 3PU, United
Kingdom.}

\date{\today}

\begin{abstract}
We present high resolution high energy and resonant x-ray diffraction results from \bilayer\ for $x=0.55,0.575$ and $0.60$.  These compounds show superlattice reflections at wavevectors of $(h\pm\delta,k\pm\delta,l)$ and
$(h\pm2\delta,k\pm2\delta,l)$, arising from orbital ordering with associated Jahn-Teller distortions and charge ordering respectively. We observe a phase transition between
the $x=0.55$ and $x=0.575$ doping levels. Samples with $x=0.55$
display structural characteristics similar to those previously reported for
$x=0.5$. Compared to this, the long range order in samples with $x=0.55$ and $x=0.6$ have a distinct change in wavevector and correlation.  We attribute this to a new orbital bi-stripe phase, accompanied by weak, frustrated, charge ordering. The observed azimuthal dependence of the orbital order reflections supports the model proposed for this new phase.
\end{abstract}

\maketitle

\section{\label{Introduction}Introduction}

\begin{figure}
\includegraphics[width=0.6\columnwidth]{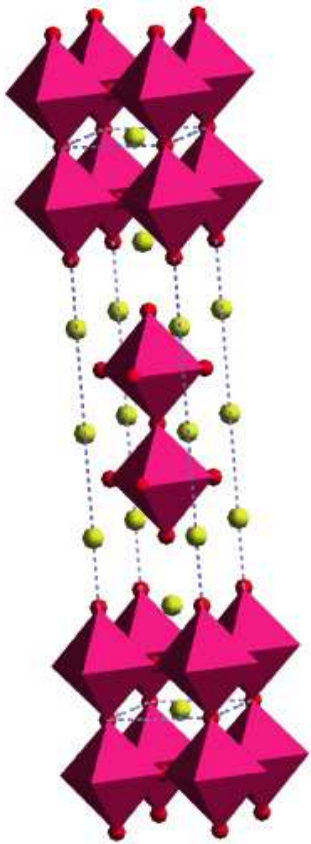}
\caption{\label{fig:crystal}(color online) Crystal structure of \bilayer. Purple octahedra respresent MnO$_6$ octahedra, yellow spheres represent La/Sr sites, and red spheres represent oxygen.}
\end{figure}

The $n=2$ member of the Ruddleston-Popper family of manganites
forms a bilayer crystal with the general formula \bilayer. The
crystal forms a layered structure consisting of two MnO layers
separated by a rock-salt type layer of (La,Sr)O. The result of
this layering is an extremely two-dimensional crystal with
$a=b=3.87~\textrm{\AA}$ and $c=19.95~\textrm{\AA}$ (fig~\ref{fig:crystal}). 

\begin{figure}
\includegraphics[width=\columnwidth]{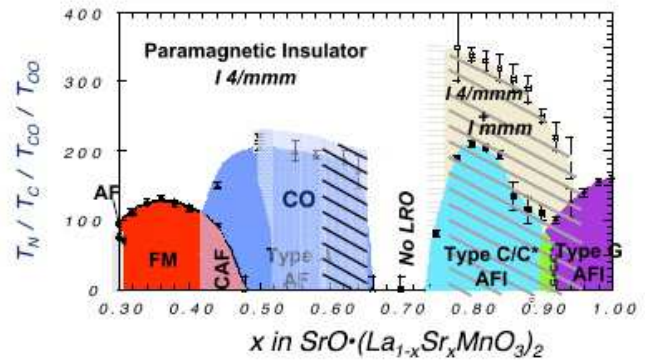}
\caption{\label{fig:qui}(color online). Structural and phase diagram from Qui \emph{et al.}\cite{qui} of the bilayer manganite.  The region of interest in this paper is the area with diagonal black stripes around $x=0.6$.   In this paper we propose a different orbital and charge order to that of the half doped bilayer.}
\end{figure}

The \bilayer\ system presents a very complicated phase diagram.  The discovery of  colossal magneto-resistance in the $x=0.4$ bilayer\cite{moritomo:380} triggered a flurry of interest in this compound\cite{koizumi:060401}.   Following this, numerous studies have been conducted on the $x=0.5$ doping level, showing strong charge and orbital order order. Far less work has been done on the overdoped crystals, however the work that has been done shows many interesting effects.   Neutron studies reported that there exists a number of distinct magnetic phases\cite{ling:15096} throughout the stochiometric range (fig.~\ref{fig:qui}).   In the area surrounding the half doped region $0.46\leq x\leq0.66$ the low temperature ordering is in a type-A antiferromagnetic phase.   Above this there appears to be a gap where there exists no long range magnetic or charge order.  This is unique to the bilayer system, and is not observed either in the single layered, or cubic, manganites.   Above $x=0.74$ long-range ordering is reformed, but this time in a type-C/C* magnetic phase.    Finally, above $x=0.9$ the system enters a type-G phase.

In this paper we will show data from the upper end of the type-A phase with $0.55<x<0.6$.   We have applied resonant and high energy x-ray diffraction to look at the superlattice peaks associated with the charge and orbital ordering as well as the structural distortions in the compounds with $x=0.55,0.575$ and $x=0.6$ doping levels.   We observe a distinct change between the $x=0.55$ and $x=0.575$ doping levels, characterised by an increase in the phase transition temperature associated with both charge ordering and concomitant Jahn-Teller structural distortions (T$_{CO/JT}$) coupled with a discontinuous change in the ordering wavevector.   We also report a strong correlation between the Jahn-Teller distortion peak intensity and the commensurability of the superlattice reflections in the $x=0.575$ and $0.6$ samples.   The wavevector of the orbital order at low temperature suggests a periodicity five times larger than that of the chemical unit cell in the \emph{ab} plane.   We propose a quasi-bistripe phase of orbital order, complemented by a weak frustrated charge ordering.  A charge and orbital order pattern has been constructed, and is proposed as the structure in the range $0.55<x<0.625$. Such a structure would have a particular polarisation and azimuthal dependence of the orbital order reflection which has been simulated.  Our experimental results at the Mn $K$ edge confirm the predicted azimuthal dependence giving us confidence in the proposed structure observed using both resonant x-ray and high energy x-ray diffraction. The orbital order is accompanied with a structural Jahn-Teller distortion, which appears to be more persistent and stable than in the half doped bilayer manganites.

\section{\label{results}Experiment and Results}
High quality single crystals of \bilayer\ 
were grown using the floating zone method at the University of
Oxford \cite{boothroyd:587}. These crystals were pre-aligned using an in-house Cu \emph{K} edge rotating anode system \cite{wilkins:2666}. Experiments were performed  using high energy (100~keV) x-ray diffraction and also resonant x-ray scattering at the Mn \emph{K} edge.

High energy x-ray diffraction was performed at the BW5 beamline at HASYLAB, Hamburg.   The beamline is equipped with a wiggler and a
water cooled Cu filter to produce x-rays in the spectral range 60-150~keV.  The
analyser and monochromator were matched SiGe graded crystals to
provide a resolution matched to the sample rocking curve.  The x-ray beam had an incident beamsize of $1\times1$~mm, and an energy of 100~keV. Detection
was provided by a solid state detector, gated to remove harmonics using a single channel analyser

The samples were mounted on a the cold finger on an APD displex
cryofurnace capable of a temperature range $10$~K$ <T< 400$~K.
Orientation was such that the \emph{c} axis was parallel to the
incident beam, and the \emph{ab} plane perpendicular.

Resonant x-ray diffraction was undertaken at the XMaS UK CRG beamline at the ESRF \cite{brown:1172}.   An incident beam energy in the region of 6.555 keV (Mn $K$ edge) was provided by a double bounce Si(111) water cooled monochromator, with harmonic rejection mirrors.   Crystals pre-cleaved with the \emph{c} axis surface normal were mounted with the \emph{c} axis along the scattering vector, allowing access to $(00l)$ type reflections.   The sample environment was similar to that at BW5 with a closed circle cryostat held in a Eulerian cradle.   A Cu (220) crystal was used for polarization analysis which at 6.555 keV has a scattering vector $47^{\circ}$ from the incident beam. This allows a leakthrough of $\sim3.5\%$ between the two polarization channels.

We describe our results in the following three sections organized by doping stoichiometry, starting by the composition most similar to the well characterised $x=0.5$ composition.

\subsection {\label{x=55}$x=0.55$}
\begin{figure}
\includegraphics[width=0.8\columnwidth]{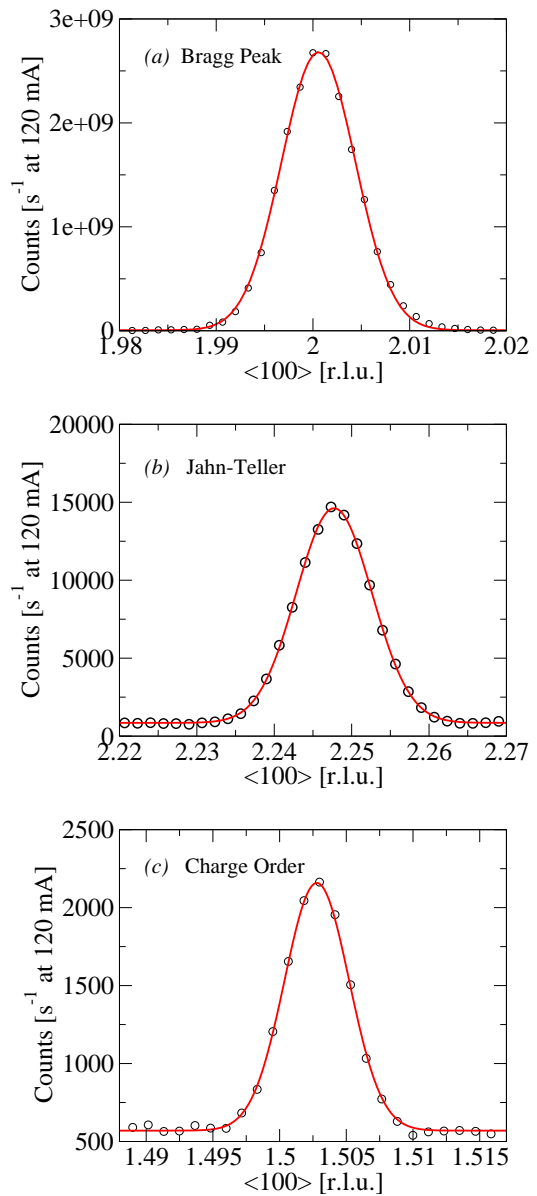}
\caption{\label{fig:peaks_55}(color online) $(2+\delta,\delta,0)$ The line shapes of the (a) (2,0,0) Bragg peak, (b) the Jahn-Teller $(2-\delta,-\delta,0)$ and (c) the charge order $(2-2\delta,-2\delta,0)$ peaks taken at 170~K measured from the $x=0.55$ sample in the [100] direction.   The solid lines show Gaussian fits to the data.   Errors are within the size of the symbols.}
\end{figure}

The sample was mounted on the high energy beamline.  Upon cooling the sample below the charge ordering temperature (T$_{CO}$) superlattice peaks appeared at wavevectors \jt. These superlattice peaks, as have
previously been described, arise from Jahn-Teller~(JT) structural
distortions. They were found regularly throughout reciprocal space
with intensities $\sim$15000~counts per second. Secondary weaker superlattice
peaks at \co, corresponding to charge ordering (CO) of the nominal Mn$^{3+}$
and Mn$^{4+}$, were also found. The observed wavevectors of both these peaks require $\delta=0.25$. These
charge order satellite peaks were $\sim$10 times weaker than those
of the JT distortions. The peak shapes from both the JT and CO signals
displayed a  Gaussian lineshape (Figure \ref{fig:peaks_55}).
This suggests that the resolution was instrument limited.
Indeed a measurement of the $(2,0,0)$ Bragg peak shows a similar
width and shape. The peaks had a far greater width in the [001] direction. This we attribute of the two-dimensional nature of the crystal structure.

\begin{figure}
\includegraphics[width=\columnwidth]{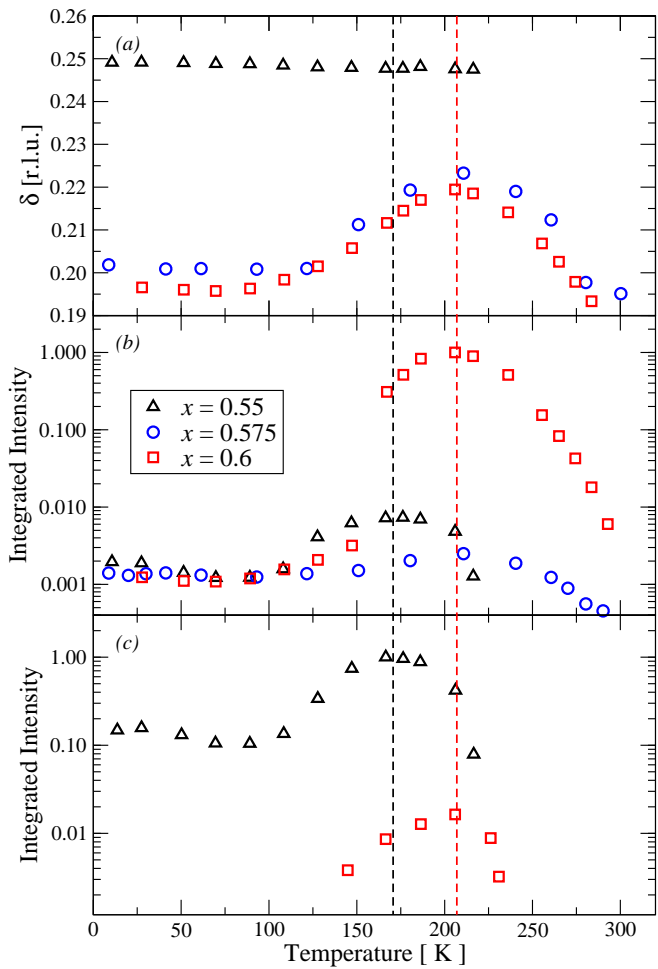}
\caption{\label{fig:int+wave}(color online) (a) \emph{top panel} Measurement of the wavevector of the Jahn Teller distortion peak at $(h\pm\delta, k\pm\delta, l)$ for the doping levels $x=0.55$ (black triangles), $x=0.575$ (blue circles), and $x=0.6$ (red squares) as a function of temperature upon warming from base temperature.  (b) \emph{middle panel} Integrated intensity of the Jahn Teller peak for the $x=0.55,0.575,0.6$ (as above).  (c) \emph{lower panel} Intensity of the charge order in the $x=0.55$ (black triangles) and $x=0.6$ (red squares) doping levels.}
\end{figure}

The $x=0.55$ sample was cooled to the base temperature of 12~K, and the
JT and CO peaks were measured upon warming. At each temperature thermal equilibrium was achieved before the intensity and the width were measured.  In order to accurately measure the commensurate wavevector, the position of two satellite peaks opposite each other with respect to a Bragg peak was determined. No significant change of the position or peak width
was detected throughout the temperature range (Figure~\ref{fig:int+wave}). The measured integrated intensity
displayed a significant increase at $\sim$120~K and then reached a
maximum at T$_N$ (180~K).  The intensity
of the peaks then fell sharply with increasing temperature, until
reaching background at 220~K. This behavior is extremely similar
to the $x=0.475$ and $0.5$ compounds. We did observe a slight increase in
the transition temperatures (T$_N$, T$_{CO}$) of about 10~K
compared to that in the $x=0.5$ sample.

\subsection{\label{x=0.575}$x=0.575$}

The only satellite peaks detected by high energy x-ray diffraction were located at $(h\pm\delta,k\pm\delta,l)$ positions. These peaks associated with JT distortions were significantly
weaker than those found in the $x=0.55$ sample. Comparing the relative intensities of the peak strength at $\pm2\delta$ with that at $\pm\delta$ in the $x=0.55$ system, a similarly proportioned signal in the $x=0.575$ sample would have been extremely difficult to detect.   As such we suspect that charge ordering does exist but it is too weak for us to detect.   The peak at $(2-\delta,-\delta,0)$ was much broader in the $x=0.575$ than in the $x=0.55$ sample, and therefore the measurements were not limited by the instrument resolution. The shape of the peak was Lorentzian squared in the high resolution [001] direction. This suggests a that the resolution effects are negligible to the width of the peak, despite this we observed no significant variation of the peak width with temperature. Unlike the $x=0.55$ sample the Jahn-Teller signal in the  $x=0.575$ sample does display a significant variation in the wavevector, $\delta$ (fig.~\ref{fig:int+wave}(a)). This variation follows a strikingly similar pattern to the variation of the intensity of the JT distortion as a function of temperature. Initially at low temperature
$\delta\approx0.2$, however on warming and with increasing intensity this value reaches $\delta=0.22$.

\subsection{\label{x=0.6}$x=0.60$}

\begin{figure}
\includegraphics[width=0.8\columnwidth]{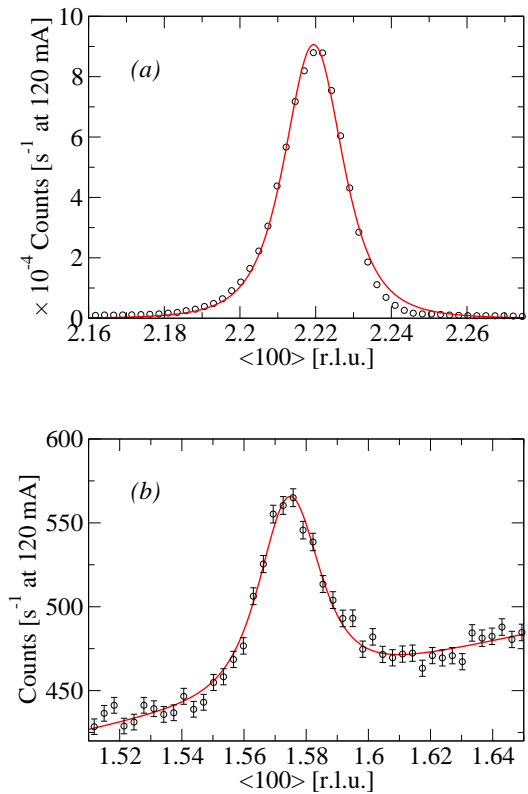}
\caption{\label{fig:peaks_60}(color online) Scans of the (a) $(2+\delta,0-\delta,0)$ Jahn-Teller
superlattice peak, and (b) the $(2-2\delta,-2\delta,0)$ charge order
superlattice peak, in the $x=0.6$ sample at 210~K measured in the [h00] direction.  Solid lines are fits to the data using Lorentzian squared
lineshapes and a linear background.}
\end{figure}

Satellite peaks were found using high energy x-ray diffraction at both $(h\pm\delta,k\pm\delta,l)$ and $(h\pm2\delta,k\pm2\delta,l)$ in the $x=0.60$ sample.   The Jahn-Teller peak is significantly stronger than that occuring in either the $x=0.55$ or $x=0.575$ doped samples.  The charge order peak however, is some 40 times lower in intensity than the Jahn-Teller peak, compared to only 10 times lower in the $x=0.55$. As with the $x=0.575$ sample the peaks are not resolution limited and they can be accurately fitted with a Lorentzian squared lineshape (Figure~\ref{fig:peaks_60}). Similar to the $x=0.575$ sample there is a significant variation of the incommensurate wavevector, $\delta$, with the intensity (fig.~\ref{fig:int+wave}), and this is present also in the charge order peaks.

Resonant diffraction at the Mn $K$ edge of the $x=0.6$ bilayer sample was performed specifically to look at the anisotropy of the structural distortion and concomitant orbital order. There has been much discussion as to what $K$ edge diffraction is sensitive to, and whether it is a direct observation of orbital ordering or not. There appears to be three possibilities.   Either the resonant signal is sensitive to the weak quadropole transition from the $1s-3d$, or to the dipole transition into the $4p$ state, which is influenced by the $3d$ level through Coulomb repulsion, or finally, that it is sensitive the dipole transition to the $4p$, which is effected by the nearest neighbour bond length and orbital occupation. The first possibility seems unlikely as one would expect a difference in energy between the quadropole and dipole magnetic signals, and as in V$_2$O$_3$\cite{paolasini:4719} it would be expected that both would be visible.  Elfimov \emph{et al.}\cite{elfimov:4264} have suggested that the Coulomb interaction proposed by Ishihara \emph{et al.}\cite{ishihara:3799} is unlikely, and the consensus seems to be that the final option is most probable. As such we are not looking at a direct probe of the orbital order, we are looking at the co-operative Jahn-Teller distortion, which accompanies the orbital ordering.  The separation of these two phenomena does appear to be possible using $L$ edge diffraction where it has been demonstrated in La$_{0.5}$Sr$_{1.5}$MnO$_4$\cite{castleton:1033,wilkins:167205}.

\begin{figure}
\includegraphics[width=\columnwidth]{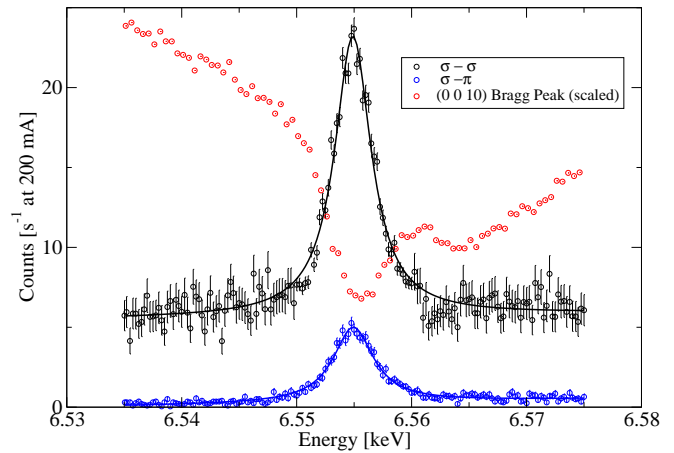}
\caption{\label{fig:energy}(color online) The energy dependence of the orbital signal at the Mn \emph{K} edge measured in the $\sigma-\sigma$ and $\sigma-\pi$ channels. Solid lines show a Lorentzian fit. The energy dependence of the $(0,0,10)$ Bragg peak for comparison with the Mn $K$ absorption edge.}
\end{figure}

The resonant signal of the  $(\delta,\delta,10)$ was collected, which was found to resonate in both the $\sigma-\sigma$ and $\sigma-\pi$ channels (fig~\ref{fig:energy}).  These resonances occurred at the same energy as the absorption edge measured at the (0,0,10) Bragg peak, and form Lorentzian lineshapes.  

The cross section for resonant scattering from an electron dipole transition (E1) can be written as follows

\begin{equation}
f^{\mathrm{xres}}_{E1}=f_{0}+if_1+f_2
\label{eqn:cross}
\end{equation}

where the terms $f_n$ are given by

\begin{eqnarray}
f_{0} & = &(\hat{\epsilon}'\cdot\hat{\epsilon})\;\; [F_{11} + F_{1-1}]\\
f_{1} & = &-(\hat{\epsilon}' \times \hat{\epsilon})\cdot \hat{z}\;\; [F_{11}-F_{1-1}]\\
f_{2} & = &(\hat{\epsilon}' \cdot \tilde{T} \cdot \hat{\epsilon})\;\; [2F_{10} - F_{11} - F_{1-1}].
\end{eqnarray}

where $\hat{\epsilon}$ and $\hat{\epsilon}'$ are the polarization vectors of the incident and scattered beam respectively,  $\hat{z}$ is a unit vector in the direction of the magnetic moment, and $\tilde{T}$ is the scattering tensor.

For $\sigma$ polarized incident light, one would only expect a signal to be present in the $\sigma-\pi$ channel, if the signal originates from the $f_1$ term.   However, the presence of $\sigma-\sigma$ scattering indicates that the resonance occurs due to terms in $f_0$ or $f_2$. As terms in $f_0$ are independent of $\vec{Q}$ we believe the resonance to occur solely from the $f_2$ term.  As expected these resonances can be fitted satisfactorily with a Lorentzian lineshape, typical of a dipole transition.   The centre of these resonances occurs at 6.555~keV corresponding to the Mn absorption edge observed of the (0,0,10) Bragg peak.   This resonant energy is identical to that seen in La$_{1-x}$Ca$_x$MnO$_3$ by XANES by Bridges \emph{et al.}\cite{bridges:R9237}, who attribute this main peak to a dipole transition.  They also see weak pre-edge features $\sim15$~eV below this which could be due either to forbidden quadropole transition or hybridization of the $4p$ level.  We do not observe these peaks by diffraction in the bilayer, however they could be within our noise level.

The azimuthal angle dependence was collected by measuring the integrated intensity of the superlattice peak in each polarization channel for a given azimuthal angle.  Due to the simultaneous presence of a signal in both polarization channels we have calculated the polarization of the scattered x-ray beam by the stokes parameter as defined by

\begin{equation}
P_1 \mathrm{\ (Stokes\ Parameter)} = \frac{I_{\sigma - \sigma }-I_{\sigma - \pi}}{I_{\sigma - \sigma }+I_{\sigma - \pi}}
\label{eqn:stokes}
\end{equation}

This has the effect of self-normalisation and removes any effect of angular changes in the size of the geometric beam footprint.  The integrated intensity of the signal in either channel was measured through a scan the polarization analyser angle $\theta$. It should be noted that the Jahn-Teller structural distortion and the orbital ordering have the same symmetry around the Mn$^{3+}$ ion.  As such this azimuthal dependence is valid for both phenomena, independant of any sensitivity arguments.

\begin{figure}
\includegraphics[width=0.8\columnwidth]{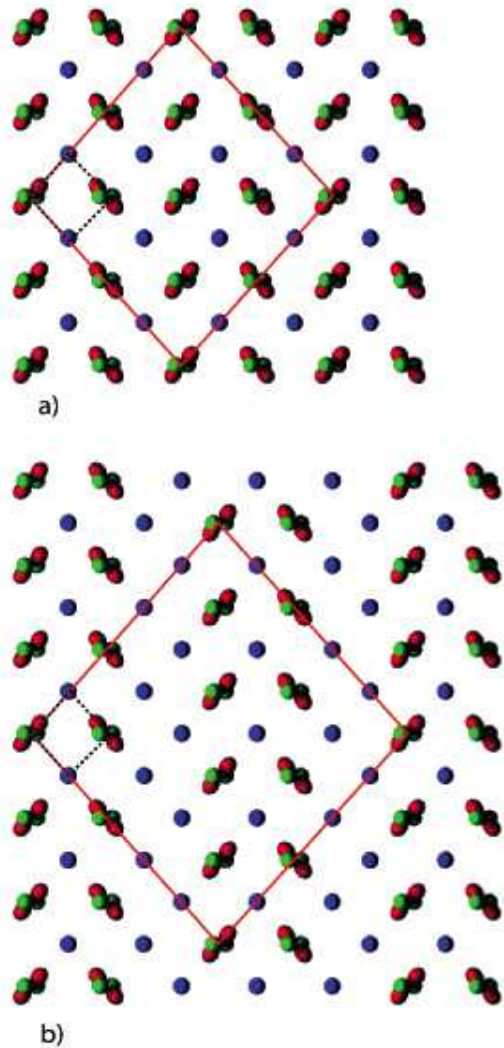}
\caption{\label{fig:cell}(color online). (a) The accepted charge and orbital ordering of \bilayer\ with $x=0.5$ (b) Proposed charge and orbital ordering for the $x=0.6$ doping level.  In both diagrams only the manganese ions are shown for clarity.  The chemical unit cell is shown by the black dotted line, and the orbital order super-cell shown by the solid red line. The orbitals are displayed as $x^2+y^2$ as these have been shown to be dominant by \emph{ab-initio} calculations\cite{koizumi:060401}.}
\end{figure}

A model of the charge and orbital order has been constructed for the high dopes phase  (fig \ref{fig:cell}). This model was made in such a way to agree with the fivefold periodicity. Using this model the azimuthal dependence has been calculated by using the Anisotropy of the Tensor of Susceptibility (ATS). On resonance the scattering on a manganese site is given by a tensor $\tilde{T}$ due to the local site symmetry $D_{4h}$\cite{carra:1509}. We therefore calculate the total structure factor from all manganese sites for the reflection (0.2,0.2,10) in the unit to obtain a single scattering tensor $\tilde{T}_{00}$.  The structure factor can then be calculated by

\begin{equation}
\tilde{T}_{00}=\sum_{i=1}^{4}\tilde{T}^{D_{4h}}_{i} e^{i\vec{Q}\cdot\vec{r}_{i}}
\label{eqn:sf}
\end{equation}

Where $\vec{Q}$ is the scattering vector and $\vec{r}_i$ are the positions of each of the four Mn$^{3+}$ ions.

The intensity can therefore be calculated using the following formula

\begin{equation}
I=|\hat{\epsilon}'\cdot\tilde{T}_{00}\cdot\hat{\epsilon}|^2
\end{equation}

\begin{figure}
\includegraphics[width=\columnwidth]{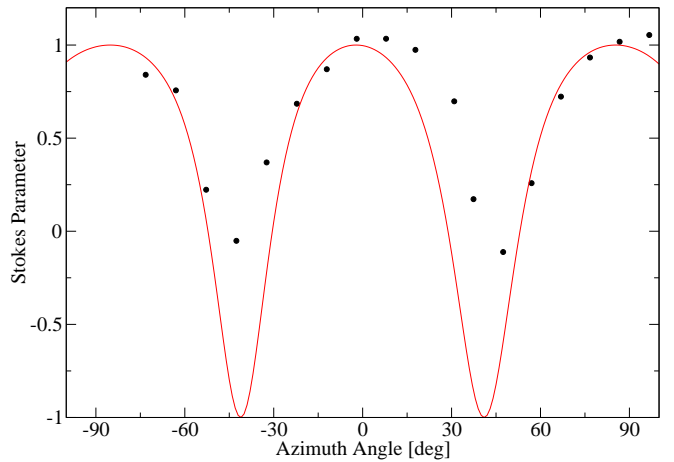}
\caption{\label{fig:azimuth}(color online) The azimuthal dependence of the Stokes parameter as calculated using the structure proposed in Fig~\ref{fig:cell} and equations \ref{eqn:stokes} and \ref{eqn:sf} (solid line) and the experimentally determined intensity of orbital order signal measured in the $\sigma-\sigma$ and $\sigma-\pi$ channels.}
\end{figure}

The azimuthal dependence of the ($\delta,\delta$,10) superlattice peak is displayed in (Fig.~\ref{fig:azimuth}), together with the results of the simulation.  The experimentally determined Stokes Parameter does not fall to -1 as predicted by the ATS simulation.  This is due to the $\sigma-\sigma$ signal being much larger than the $\sigma-\pi$ and so even a relatively small $\sigma-\sigma$ signal dramatically increases the parameter value.  This small $\sigma-\sigma$ signal could be due to small amount of background scatter present in this channel.  Overall though, there is a general agreement between the fit and the data, suggesting a correct model of the orbital anisotropy has been used.

\section{\label{discussion}Discussion}

The distinct change in the behavoir of the bilayer suggests that the higher doping, where $x\geq0.575$ (high-doped region)
forms a second `sub' phase within the charge ordered regime. Here
the JT distortion, and most probably the charge order, display
strikingly different behavior than that observed for $x\leq0.55$ (mid-doped region). The
behavior of the $x=0.55$ sample can be seen to be very similar to
the lower dopings by comparing with results presented by
Wilkins~\emph{et al.}\cite{wilkins:205110}.  The interaction between the AFM order below T$_N$ and the orbital and charge ordering appears to be unique to the bilayer manganites.  The onset of the AFM ordering simultaneously reduces the intensity of the in-plane ordering.  It was originally thought that the CO completely collapsed\cite{dho:3655} and then was re-entrant again at lower temperatures. This collapse appears to be incomplete and any increase at low temperatures is very small.   Similar behavoir was also seen in $x=0.475$ and $0.5$\cite{wilkins:205110}. It has been suggested that a spin freezing occurs below 100~K\cite{coldea:277601}, which corresponds to the minimum in the charge order reflection intensity.  This suggests that there are spin fluctuations occuring below T$_N$ which gradually slow to form a spin-frozen state at 100~K.

Unlike previous reports\cite{li:174413,larochelle:095502} we do not observe a smooth transition of the wavevector with doping level.   It was suggested that $\delta$ follows the trend $\delta=(1-x)/2$, however we observe discrete changes in the wavevector, and in particular the wavevector is not stable throughout the temperature range of the charge ordered regime. Although this trend undoubtedly seems true in general, we suggest that $\delta$ moves to the closest stable commensurate position (see table~\ref{tab:corr}).

\begin{figure}
\includegraphics[width=\columnwidth]{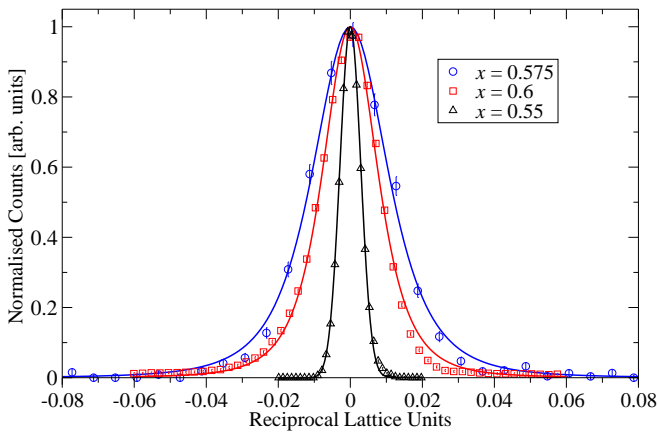}
\caption{\label{fig:jt_peaks}(color online) Comparison of the Jahn-Teller
superlattice peaks in the $x=0.55,0.575,0.60$ samples. The solid line
shows a Lorentzian squared fit for the $x=0.575$ and $0.60$ samples, and
a Gaussian fit for the $x-0.55$ sample (see section....).   The intensities of the peaks
have been normalised, and the centre of the fits set to zero.  All the linear
backgrounds have been removed after fitting the data.}
\end{figure}

The enhanced stability of the JT distorted phase was shown before by Campbell \emph{et al.}\cite{campbell:104403}, however no explanation was given for this.  We present a model showing a stable ordering of the JT distortions around the $x=0.6$ doping, however this doesn't explain why it is stable to a higher temperature than  the $x=0.5$ structure.   Indeed it is interesting that it appears that the distortions in high doping region appear to have a much shorter correlation length (see Table~\ref{tab:corr} and fig.~\ref{fig:jt_peaks}) than those in the mid doped region, and yet the superlattice peaks persevere to a high temperature. The significant change in correlation length betweeen the $x=0.55$, and the $x=0.575$ adds evidence for a phase transition between these doping levels.

\begin{table}
\caption{\label{tab:corr}Position and inverse correlation lengths
of the Jahn-Teller peaks with respect to the doping of the sample.
Inverse correlation lengths are measured at peak intensity, $\delta$ is taken at base temperature.}
\begin{ruledtabular}
\begin{tabular}{c c c}

 $x$  & Position at 10~K &  Inverse Correlation Length\\
 &($\delta$)&($10^{-2}~\textrm{\AA}^{-1}$)\\
\hline
 0.475\footnote[1]{Data taken from Wilkins \emph{et al.}\cite{wilkins:205110}} & 0.25 & 0.19 \\
 0.5\footnotemark[1] & 0.25 & 0.19 \\
 0.55  & 0.25 & $\leq$0.6\\
\hline
 0.575 & 0.20 & 2.3 \\
 0.6 & 0.20 & 1.4 \\

\end{tabular}
\end{ruledtabular}
\end{table}

The mid-doped region showed very
little change in the wavevector of the superlattice peaks, the
position of which is adequately explained using the chequerboard and CE-type model.   In the higher doped structure this symmetry breaks down,
and the positions of the superlattice peaks align at
incommensurate values.   In addition, these values change as the
intensity changes with temperature. This direct relation between
the propagation vector of the superlattice cell and the amplitude
of that cell has not been seen to this extent in other manganite.   It suggests a dynamic stripe system that
changes configuration as the degree of charge segregation alters.
In order to understand this ordered system, we first have to
understand the origin of the low temperature propagation
wavevector.

The high doped charge order phase, seems to be less correlated
than the mid dopings.   The inverse correlation length of both the
JT and charge ordering (calculated through the width of the
superlattice peaks) remain constant at all temperatures, for each
of the samples. The superlattice peak in the $x=0.55$ sample  had
a maximum inverse correlation length of
$\zeta\leq6.3\times10^{-3}~\textrm{\AA}^{-1}$, whereas the peaks from
$x=0.575$ and $x=0.6$ doped samples were
$\zeta=2.3\times10^{-2}~\textrm{\AA}^{-1}$ and
$\zeta=1.4\times10^{-2}~\textrm{\AA}^{-1}$ respectively. This
difference is clearly shown in figure~\ref{fig:jt_peaks}, where
the $x=0.55$ sample JT peak is fitted to a Gaussian lineshape,
and the much broader JT peaks from the high doped phase are
fitted to Lorentzian squared lineshapes.

Substantial discussion was generated after the original discovery
of the charge ordered systems in the \emph{AB}MnO$_{3}$ compounds,
as to the relative merits of the bi-stripe\cite{mori:473} (BS) and
Wigner\cite{radaelli:14440} crystal (WC) models\cite{khomskii:134401,hotta:4922,brey:127202}. These models provided
solutions to the orbital ordering of compounds where the ratio of nominally
Mn$^{3+}$ and Mn$^{4+}$ is not 1:1.  These same arguments can be
applied to the bilayer crystal.   Here it appears that we have an
ordering that lies midway between these models and the
checkerboard pattern. The lowest stable BS and WC models have
$x=0.66$, with an orbital orbit propagation vector
$(2\pi/a)(\frac{1}{3},0,0)$. Here we have a doping level less than
this, and the position of the superlattice peaks displays a larger propagation
vector. A satisfactory model of this mid-point is displayed in
figure~\ref{fig:cell}. This 2-1 stripe model has similarities to
both models mentioned above.  The slipage distance between successive Mn$^{3+}$ ions
alters, as one see in the BS model, however the
correlation in these species in the [110] direction, is similar to
the WC system.  The opposition of the direction of the orbitals of
the stripe pairs is necessary for the observed wavevector from the
JT superlattice.  This can be understood as simple exchange of the
orbitals between a single Mn$^{4+}$ and superexchange between two
nominal Mn$^{4+}$ ions.

Increasing the temperature of the sample alters both the intensity
and the incommensurate state of the charge ordered peaks of the
high doped phase.  The wavevector $\delta$ increases, however does not reach the
stable mid-doping $\delta=0.25$.   As $\delta$ increases we can
imagine the 2-1 stripe model gradually turning into the mid-doped
stripe phase by losing double stripes.  The incommensurability
reaches a point of maximum intensity where every other double
stripe is now a single stripe and so instead of a double stripe
and then a single stripe repeated, there is now a double stripe
and then two single stripes before a second double stripe.  It
would be expected that this model would not be well correlated as
there are not sufficient Mn$^{3+}$ ions.  As the temperature increases further, and the
intensity decreases, $\delta$ falls back to the $\delta=0.2$ high
doped phase value.

The measurements of the azimuthal dependence of the orbital order is a method of testing the order model.  Different orbital patterns result in a different azimuthal dependence and the strong agreement between calculated and measured values corresponds to this model. A similar study has been reported by Di Matteo \emph{et al.}\cite{matteo:024414}, on the $x=0.5$ doped bilayer.  In their paper they simulated the traditional Jahn-Teller distorted checkerboard pattern and found an excellent agreement with their experimental data.

\section{\label{conc}Conclusions}

We have presented results from high resolution x-ray scattering studies of La$_{1+2x}$Sr$_{2-2x}$Mn$_2$0$_7$
where $x=0.55$, $575$ and $0.6$. It is clearly demonstrated that there is a
distinct change in the nature of the charge ordering and
accompanying Jahn-Teller distortion  at $0.55<x<0.575$  The
$x=0.55$ sample shows very similar behavior to that seen in
$x=0.5$, whereas the $x\geq0.575$ samples show ordering with a
much lower correlation. A striking incommensurate behaviour is also
seen in this higher doped charge ordered phase. A new model containing quasi-bistripe ordering is proposed, and the measured azimuthal dependence of the orbital order agrees with this model.  We suggest that this quasi-bistripe ordering would turn into a true bistripe order as the doping level is increased further towards $x=0.66$.

\section{\label{acknow}Acknowledgements}
TAWB and PDS wish to thank EPSRC for support. PDH
thanks the University of Durham Research Foundation for support. SBW would would like to thank the European Commission for the support in the frame of
the ``Training and Mobility of Researchers'' program.

\bibliography{bilayer}

\end{document}